\documentstyle[prd,aps]{revtex}
\begin{document}
\draft
\twocolumn[\hsize\textwidth\columnwidth\hsize\csname
@twocolumnfalse\endcsname
 \preprint{SU-ITP-96-53,  hep-th/9611162}
\title{ Bound States of Branes with Minimal Energy}
\author{Renata Kallosh}
\address{Physics Department, Stanford University,
Stanford CA 94305-4060, USA}
\date{November 20, 1996}
\maketitle
\begin{abstract}
It is pointed out that  the energy of the bound states of D-branes and strings
is determined by  the central charge of the space-time supersymmetry. The
universality which is seen at the black hole horizon appears  also on the
D-brane side:  the total energy of the bound states of a given number of branes
has a minimum  when considered as a function of  the independent parameters
(moduli). This provides a new evidence that the near-horizon space-time
geometry  of the dilaton black holes can be represented by the bound states of
branes. The axion-dilaton dyonic black holes have the mass formula of the
non-threshold type bound state.  Upon uplifting to higher dimensions they may
give information about such states.
\end{abstract}
\pacs{PACS: 11.25.-w, 04.65.+e, 04.70.Dy, 11.30.Pb  \hspace{1.4 cm}
SU-ITP-96-53
\hspace {1.4 cm}
hep-th/9611162}
\vskip2pc]

 Supersymmetric General Relativity (SGR)  and Quantum Mechanics (QM) have been
recently related to each other \cite{VS}  using the D-brane technology
\cite{Pol}, see \cite{Mald} for  a review on black holes in string theory. In
SGR one of the main features is the universality property of supersymmetric
black holes near the horizon. It reveals in the fact that the 4- and
5-dimensional black holes are attractor systems with fixed points \cite{FKS}
for moduli. The area of the horizon is universally defined by the central
charge, extremized in the moduli space \cite{FK}.

On the other hand, the QM side of the picture so far has not shown any signs of
the universality even  for the particular black holes for which the geometry of
 space-time near the horizon  is replaced by   D-branes and strings.

Here we would like to show how the universality property of supersymmetric
black holes near the horizon can be seen on the QM side.  Our conjecture will
be general, but we will be able to check it only for limited examples
available.
Consider the supersymmetry algebra
\begin{equation}
\{Q_\alpha, Q_\beta\} =S_{\alpha\beta}
\label{eq:onea}
\end{equation}
in some dimension. Examples in D = 12, 11, 10 and 4  relevant to   this
discussion can be found in the literature, see \cite{bars} for the most recent
discussion from the point of view of d=12 and \cite{Pol} in the context of
D-branes.  The right-hand side  of the  algebra contains translation as well as
all possible p-form charges  permitted by symmetry \cite{townsend,bars,Pol}.
The state with some unbroken supersymmetry is characterized by a vanishing
eigenvalue, or eigenvalues of this matrix \cite{K,kal-raj,bars}
\begin{equation}
\det  \left[S_{\alpha\beta} \right] \mid BPS> =0 \ .
\end{equation}
We will consider here only   massive states, which means that there exists a
rest frame where the energy of the state which is equal to its mass is given by
(in the basis
where the eigenvalues of the central charge matrix are real)
\begin{equation}
tr \left[S_{\alpha\beta} \right] \mid BPS> = H \mid BPS> = E \mid BPS> \ .
\end{equation}
The SGR part of universality in 4 and 5 dimensions comes from the observation
that the central charge  in the SUSY algebra is defined by the charge of the
graviphoton, which is a particular combination of moduli and quantized charges.
 Supersymmetric theories with $N>2$ which have few eigenvalues of the central
charge matrix  can be understood in terms of N=2 theories.  The universality is
in the fact that near the horizon the largest eigenvalue of the central charge
is extremized in the moduli space, which requires all other eigenvalues to
vanish at the attractor.  The entropy $S$ of the black hole is given by the
volume of the $S^2$ or $S^3$ sphere of the radius $r_h$   in d=4 and d=5
respectively \cite{FK}:
\begin{eqnarray}
S(d=4) &=& {A\over 4} = \pi (r_h)^2   = \pi (Z \bar Z )_{\rm fix} \ ,  \\
S(d=5) &=& {A\over 4} = {\pi^2\over 2}  (r_h)^3   ={\pi^2\over 2}  Z_{\rm
fix}^{3/2} \ .
\end{eqnarray}
Even more interesting for our purpose here is the fact that simultaneously with
defining the entropy, the extremized value of the central charge defines the
extremized value of the ADM mass of the black hole \cite{FK}:
\begin{eqnarray}
M_{ADM}^{\rm min} (d=4) &=& \sqrt {(Z \bar Z )_{\rm fix} } \ , \\
  M_{ADM}^{\rm min}(d=5)  &=& {3\pi\over 4}  Z_{\rm fix} \ .
\end{eqnarray}
The central charge is defined by supergravity theory under consideration and
depends on moduli $\phi$ and quantized charges $(p,q)$. In d=4 the central
charge, as well as the moduli, is complex, and in d=5 both are real.
Upon extremization moduli become fixed functions of charges $(p,q)$ and the
entropy as well as the extremized value of the ADM mass of the black hole
depend only on charges. It has been proved recently that in N=2 theories with
special geometry the extremal value of the supersymmetric black hole mass is
the minimal one \cite{FGK}.
\begin{equation}
|Z(\phi, (q,p) ) |\geq |Z(\phi, (q,p) ) |_{{\partial |Z| \over \partial
\phi}=0} = |Z(\phi_{\rm fix} (q,p), (q,p) )| \ .
\end{equation}
The simplest way to understand how many moduli are fixed at the black hole
horizon is to perform a consistent truncation to N=2 theory, starting with
N=8,4. Some moduli go into vector multiplets, some into hypermultiplets. Only
the first ones are fixed in terms of charges, the scalars of the
hypermultiplets are not fixed  in terms of charges although in all known
supersymmetric black hole solutions they are constants \cite{FK}. The number of
scalars from the vector multiplets equals to the number of gauge groups minus
one since one vector belongs to the gravitational multiplet. In d=4 each gauge
group may contain both electric and magnetic charge and each scalar may be
complex. In d=5 there are only electric charges in each gauge group and the
scalars are real.

Superstring  theory  supplies
the supersymmetry generators $Q_\alpha$ build out from the quantum mechanical
states of a string and D-branes. One may try to find the relation between the
universal picture in SGR  and
those examples  where the black holes are understood at the QM level.

One may  expect that  when only superstring excitations, including the
D-branes, are taken into account, they may not be sufficient for the complete
universal picture at the QM side. One may need to include some QM excitations
related to the 2- and 5-branes of the 11-dimensional theory or some other
modes. This  is indeed the case studied for toroidal compactification in
\cite{KlebTsey} and for more general Calabi-Yau black holes  in
\cite{klausthomas}.

Our main  conjecture is  that in general the QM interpretation of the universal
minimization of the central charge is the following.
The energy of the bound  states of various branes depends on the number  of
branes and on their energies. The energy-charge of the individual brane states
are functions of some independent parameters, which can be identified as
moduli, see examples below. The  value of the total energy of a given number of
brane states is minimal
when the energies of the branes pick up the values depending on the number of
branes of all kinds, making the bound state with the minimal energy.

Thus we would like to show that the energy of the BPS bound state $E_{\rm bs}$
of various number of branes has the energy which reaches the minimum when the
energies of the individual branes $e_{\rm br}$ become particular functions of
the numbers of branes $n_{\rm br}$.
\begin{equation}
E_{\rm bs} \left (e_{\rm br} (\phi), n_{\rm br}\right )  \geq  E_{\rm bs}^{\rm
min} (e_{\rm br} \left(\phi (n_{\rm br})\right) , n_{\rm br})  \ .
\end{equation}
The minimization condition is
\begin{equation}
{\partial E_{\rm bs} \left (e_{\rm br} (\phi), n_{\rm br}\right )\over
\partial \phi } =0  \ .
\end{equation}
The minimal value of the energy of the bound state correspond exactly to the
ADM mass of the double-extreme black holes with constant moduli \cite{KSW} when
the number of branes is identified with the quantized charges of the black
holes.
Consider the first example  \cite{HMS}, related to 5d black hole: the bound
state of some number   $Q_1$ of D-onebranes,   $Q_5$ of D-fivebranes,  and
Kaluza-Klein momentum in type IIB string theory compactified on $T^5$. The
D-fivebranes are wrapped on $T^5$. $Q_1$ D-strings are wrapped along the
direction 9 and there is a momentum $P_9= {N\over R^9}$ along the string which
is in direction 9.

Here the 10d Newton constant and 5d Newton constant are related as follows:
$$G_{N}^{10} = 8\pi^2 g^2 = G_{N}^5  (2\pi)^5 R_5 R_6 R_7 R_8 R_9= G_{N}^5
(2\pi)^5 V R_9 \ .$$
We are interested in the QM of the near horizon 5d black holes. Therefore in
what follows we will fix the 5d Newton constant $G_{N}^5$, whereas the 10d
Newton constant $G_{N}^{10}= 8\pi^2 g^2$ will be subject of variation. As long
as the radius of 9-th direction $R_9$ and the string coupling constant are
arbitrary, the energy of the BPS bound state described above is (in $\alpha'=1$
units)
\begin{equation}
E= {R_9 |Q_1|\over g} + {R_9 V |Q_5|\over g} +{ |N|\over R_9} = E_1 + E_5 +E_P
\ .
\label{energy}\end{equation}
In this example the energy of the bound state is the sum of the energies of the
constituents.
We are interested in the variation of this energy at the fixed values of the
number  $Q_1$ of D-onebranes,  $Q_5$ of D-fivebranes and string states $N$ as
well as at the fixed $G_{N}^5$. To perform this variation we rewrite eq.
(\ref{energy}) as
\begin{equation}
E(g, R_9) = {R_9 |Q_1|\over g} + { g|Q_5|\over 4 \pi^3 G_N^5} +{ |N|\over R_9}
\ .
\end{equation}
For simplicity we will fix  $4 \pi^3 G_N^5=1$.
We proceed with extremization of the energy as a function of $g$ and $ R_9$,
starting with
\begin{eqnarray}
E(g, R_9) &=& {R_9 \over g} |Q_1|+  g|Q_5| +{ 1\over R_9} |N| \nonumber\\
&=&e_1|Q_1|+e_5|Q_5|+e_P|N| \ .
\end{eqnarray}
Variation of the total energy over the string coupling constant gives
\begin{equation}
{\partial E \over \partial g} = - {R_9 |Q_1|\over g^2} +  |Q_5|=0 \ .
\end{equation}
Variation of the total energy over the radius of the 9-th dimension leads to
\begin{equation}
{\partial E \over \partial R_9} =  { |Q_1|\over g} - { |N|\over R_9^2}  =0 \ .
\end{equation}
The minimum of the energy of our bound state is reached when the string
coupling constant and the radius of the 9-th dimension take the fixed values,
prescribed by the integer numbers $Q_1, Q_5, N$:
\begin{eqnarray}
g_{\rm fix} &=&\left \{{ |Q_1 Q_5 N|\over |Q_5|^3 }\right \}^{1/3} \ ,
\nonumber\\
\nonumber \\
  \left({1\over  R_9 } \right)_{\rm fix} &=& \left \{{ |Q_1 Q_5 N|\over |N|^3
}\right \}^{1/3}  \ ,
\end{eqnarray}
and as a consequence,
\begin{equation}
\left({R_9 \over  g}\right)_{\rm fix} = \left \{{ |Q_1 Q_5 N|\over |Q_1|^3
}\right \}^{1/3} \ .
\end{equation}

The energy of the bound state at the fixed values of the string coupling
constant and the radius of the 9-th dimension in units we have chosen becomes
\begin{equation}
E^{\rm min} (g_{\rm fix}, (R_9)_{\rm fix} ) = 3  |Q_1 Q_5 N| ^{1/3} \ .
\end{equation}
It consists of 3 equal contributions: from all    D-onebranes, from all
D-fivebranes and from all string states:
\begin{equation}
 E^{\rm min} = 3 \left({R_9 \over  g}\right)_{\rm fix} |Q_1| =  3  g_{\rm fix}
|Q_5| = 3 { |N|\over (R_9)_{\rm fix}} \ .
\end{equation}

The black hole  entropy  is proportional to the minimal energy of the bound
state in power $3/2$ as expected.
Thus in this example we have shown that the minimal energy of the bound state
is topological at the fixed values of $G_N^5$. This is a QM interpretation of
the minimization of the black hole area and the black hole ADM mass.

One can rewrite this construction which describes the bound state of D-branes
and string states in very special geometry terms adapted to 5d black holes
\cite{FK} where we identify
\begin{eqnarray}
&&E= Z= t^I(\phi_1, \phi_2)  q_I = t^0 q_0 + t^1 q_1 + t^2 q_2, \nonumber \\
\nonumber \\
 && t^0 t^1 t^2 =1 ,\\
\nonumber\\
 &&Q_1=q_0, \;
Q_5=q_1, \; N=q_2, \ \phi_1 = g, \; \phi_2 = R_9 . \nonumber
\end{eqnarray}
The minimum of the central charge now is identified with the minimum of the
bound state energy:
\begin{equation}
Z_{\rm fix} = E^{\rm min} (g_{\rm fix}, R_{\rm fix} ) = 3  |q_0q_1q_2| ^{1/3} =
3  |Q_1 Q_5 N| ^{1/3} \ .
\end{equation}
The very special geometry construction is easily generalizable to more general
cases when $d_{IJK} t^I t^J t^K =1$,   $I=0,1,\dots , n.$ The QM side still has
to be found.

Consider the second example  \cite{MS}, related to 4d black hole:
the bound state of some number   $Q_2$ of D-twobranes, $Q_5$ of solitonic
fivebranes,   $Q_6$ of D-sixbranes and Kaluza-Klein momentum in type IIB string
theory compactified on $T^6$.  Here the procedure is practically  the same as
in 5d case.
One starts with the energy of the bound state consisting of the sum of the
energy of the constituents.
\begin{eqnarray}
E&=& {R_9 R_4 \over g} |Q_2|\ + {R_9 V \over g^2 } |Q_5|+{R_9 R_4 V \over g }
|Q_6|+{ 1\over R_1} |N |\nonumber\\
&=& E_2 + E_5 +E_6 + E_P \ .
\label{energy4}\end{eqnarray}
Here $V=R_5R_6R_7R_8$ as before, $R_9$ is the radius  of the 9-th dimension,
and $R_4$ is the radius of the 4-th dimension. We  fix  the 4d Newton constant
$8 G_N^4 = {g^2 \over R_4 V R_9} =1$ for simplicity and the bound state energy
is:
\begin{eqnarray}
E&=& {R_9 R_4 \over g} |Q_2|\ + {1 \over R_4 } |Q_5|+g  |Q_6|+{ 1\over R_9} |N
|\nonumber\\ &=& E_2 + E_5 +E_6 + E_P \ .
\label{energysimpl}\end{eqnarray}
Varying it over $R_9, R_4,g$ we get
\begin{eqnarray}
&&{ R_4 \over g} |Q_2|\ -{ 1\over R_9^2} |N |=0 \ , \\
&&{R_9  \over g} |Q_2|\ - {1 \over R_4^2 } |Q_5|=0 \ , \\
&&{R_9 R_4 \over g^2} |Q_2|-   |Q_6|=0 \ .
\end{eqnarray}
The solution for the fixed values of string coupling and dimensions of two
circles  is
\begin{eqnarray}
g_{\rm fix} &=&\left \{{ |Q_2 Q_5 Q_6 N|\over |Q_6|^4 }\right \}^{1/4}\ ,
\nonumber\\
  {1\over  (R_4)_{\rm fix}} &=& \left \{{ |Q_2 Q_5 Q_6 N|\over |Q_5|^4 }\right
\}^{1/4}\ ,\nonumber\\
{1\over  (R_9)_{\rm fix}} &=& \left \{{ |Q_2 Q_5 Q_6 N|\over |N|^4 }\right
\}^{1/4} \ .
\end{eqnarray}
Again the useful derivable object is
\begin{equation}
 \left ({R_9 R_4 \over  g} \right)_{\rm fix}=
 \left \{{ |Q_2 Q_5 Q_6 N|\over |Q_2|^4 }\right \}^{1/4} \ .
\end{equation}
The minimal energy of the bound state is achieved when the contribution from
the
D-twobranes,  solitonic fivebranes,  D-sixbranes and  string states equal each
other:
\begin{eqnarray}
E^{\rm min} &=& 4 \left({R_9 R_4 \over g}\right)_{\rm fix}  |Q_2| = 4 \left( {1
\over R_4 }\right)_{\rm fix} |Q_5|\nonumber\\
&=& 4 g_{\rm fix}  |Q_6|=4 \left({ 1\over R_9}\right)_{\rm fix} |N|
 \ ,\label{energyequal}\end{eqnarray}
and \begin{equation}
E^{\rm min} = 4 |Q_2 Q_5 Q_6 N|^{1/4} \ .
\label{energymin}\end{equation}
Relation to the entropy is simple\footnote{In deriving this relation in the
standard form we had to take into account that  our units  with $8 G_N^4 =1$,
which we took for simplicity, has to be changed which provides
the correct relation.}:
\begin{equation}
S= \pi |Z_{\rm fix}|^2 = \pi (E^{\rm min})^2 = 2\pi  |Q_2 Q_5 Q_6 N|^{1/2} \ .
\end{equation}
Note that in more general situation when we have 4d black holes of N=2
supergravity
with $n$ vector multiplets the central charge is complex \cite{Cer,FK} and the
formula for the energy of the bound state is not given by the sum of the
energies of the constituents
\begin{equation}
E=\sum_i E_i\ .
\end{equation}
In general the energy of the bound state of some QM system (which has to
replace the near horizon black hole) is defined by a quadratic relation
\begin{eqnarray}
E^2 (\phi, \bar \phi, p,q)  &=&  \sum_{IJ} e^{K( \phi, \bar \phi) }  \Bigl(
F_I (\phi) \bar F_J(\bar \phi)  p^I p^J ~~~~\nonumber\\
 &+&X^I (\phi) \bar X^J(\bar \phi)  q_I q_J
- F_I (\phi)    \bar X^J(\bar \phi) p^I  q_J \nonumber\\
\nonumber\\
 & -& X^I (\phi) \bar F_J (\bar \phi)\ q_I p^J \Bigr),
\label{non-thr}\end{eqnarray}
and the minimal energy is given when the moduli take the fixed values
 $\phi_{\rm fix}(p,q)$ defined by extremization of the energy.
\begin{equation}
E^{\rm min} ( p,q) = E (\phi, \bar \phi, p,q)_{\rm fix}\ .
\end{equation}
Only in special cases when the central charge is real the energy is given by a
sum of the energies of constituents whose numbers are $p^I$ or $q_I$, and the
number of moduli is $n$:
\begin{equation}
E   = \sum _{I=0} ^{n} e^{K(\phi) \over 2 }  \left( X^I (\phi)  q_I  -  F_I
(\phi)  p^I \right )\ .
\end{equation}
The condition when this happens is that in a given gauge group either $p^I$ or
$q_I$ are vanishing. If at least in one gauge group both $q$ and $p$ do not
vanish, the
energy is not the sum of the energies of constituents.
All examples of black holes-D-branes above obviously fall into the class of
bound states with sum of energies, or so called threshold bound states. Indeed
they are dilaton-no-axion solutions. The same is true for the CY no-axion black
holes of \cite{BCdWKLM}  which have been recently understood by counting open
2-branes which end on the M-5-brane in \cite{klausthomas}.

The non-threshold bound states eventually will be identified with the dyonic
black holes which have electric and magnetic charges in some gauge groups and
complex moduli, i.e.  both dilatons as well as axions are present in the
solutions.  Many such black hole solutions are known in SGR, see
\cite{axion,BCdWKLM}. The simplest one is the axion-dilaton $SL(2,Z)$-invariant
 solution \cite{axion}.
The duality invariant mass formula for these dyonic black holes is  always of
the type shown in eq. (\ref{non-thr}).
It remains to find their QM description. One may expect that the uplifting of
such black holes to d=10, 11 will explain the structure of such bound states,
where the energy of the total state is not a sum of energies of the
constituents. It is, however, guaranteed that for the regular black holes with
the singularity covered by the horizon  the  energy will have a minimum,
related to the axion-dilaton black hole entropy.

\

I appreciate useful discussions with E. Halyo, B. Kol, A. Linde, A. Rajaraman
and L. Susskind. This work is  supported
by  NSF grant PHY-9219345.

\newpage

\end{document}